\documentclass[graybox]{svmult}

\usepackage{amsfonts}
\usepackage{amssymb}
\usepackage{amstext}

\usepackage{mathptmx}        
\usepackage{helvet}          
\usepackage{courier}         
\usepackage{type1cm}         
\usepackage{makeidx}         
\usepackage{graphicx}        
\usepackage{multicol}        
\usepackage[bottom]{footmisc}

\def\be{\begin{equation}}
\def\ee{\end{equation}}
\def\ba{\begin{eqnarray}}
\def\ea{\end{eqnarray}}
\def\lb{\label}
\def\nn{\nonumber}

\def\a{\alpha}
\def\b{\beta}

\def\d{\delta}

\def\e{\varepsilon}
\def\l{\lambda}

\def\s{\sigma}

\def\L{\Lambda}

\def\id{\mbox{\em 1\hspace{-3.4pt}I}}
\def\subbbc{{\rm C}\kern-3.3pt\hbox{\vrule height4.8pt width0.4pt}\,}

\def\vac{\mid 0 \rangle}

\begin{document}

\title*{On quantum WZNW monodromy matrix -- factorization, diagonalization, and determinant}

\titlerunning{On WZNW quantum monodromy matrix}

\author{Ludmil Hadjiivanov and Paolo Furlan}

\institute{Ludmil Hadjiivanov \at Theoretical and Mathematical Physics Division,
Institute for Nuclear Research and Nuclear Energy,
Bulgarian Academy of Sciences, Tsarigradsko Chaussee 72, BG-1784 Sofia, Bulgaria and
INFN, Sezione di Trieste, Trieste, Italy,
\email{lhadji@inrne.bas.bg}
\and Paolo Furlan \at Dipartimento di Fisica dell' Universit\`a degli Studi di Trieste,
Strada Costiera 11, I-34014 Trieste, Italy and
INFN, Sezione di Trieste, Trieste, Italy,
\email{furlan@trieste.infn.it}}

\maketitle

\abstract{We review the basic algebraic properties of the quantum monodromy matrix $M$ in the
canonically quantized chiral $SU(n)_k$ Wess-Zumino-Novikov-Witten model with a quantum group symmetry.}

\section{Introduction}
\label{sec:1}

The Wess-Zumino-Novikov-Witten (WZNW) model \cite{W} on a $2D$ cylindric space-time (with periodic space coordinate)
describes the conformal invariant free motion of a closed string on a Lie group manifold \cite{GW}.
We will only consider here the case of a compact semisimple Lie group $G\,$ and positive integer level $k\,,$
and the explicit calculations will apply exclusively to $G=SU(n)$. Canonical quantization prescribes
replacing the classical Poisson brackets (PB) by commutators or, in the case of quadratic PB, by {\em exchange relations} such
that the classical symmetries are recovered in the quasiclassical limit. Here is a short list of references on the subject covered below:
\cite{F1, AS, G, FG1, BFP, HIOPT, FHIOPT, FH2}.

The $2D$ WZNW field admits a chiral splitting in a product of left and right movers.
The chiral field $g(z)$ (where $z=e^{ix}\,$ and $x\,$ is a light cone variable) is only twisted-periodic,
\be
g(e^{2\pi i} z) = g(z) \, M\ ,
\lb{gM}
\ee
where $M\,$ is the {\em monodromy matrix}\footnote{We start with a general monodromy matrix (clasically,
$M\in G$). The case when $M$ belongs to the maximal torus will be considered later as a diagonalization problem.
The possibility of analytic continuation in $z\,$ (in correlation functions) due to energy positivity is
implicitly assumed.}.
The corresponding exchange relations with a constant statistics matrix $\hat R$ read
\be
g^A_{~\a}(z_1)\, g^B_{~\b} (z_2) =\,
\stackrel{\curvearrowright}{g^B_{~\rho} (z_2)\, g^A_{~\s} (z_1)}\!{\hat R}^{\rho\s}_{~\a\b}\qquad
(\,|z_1| > |z_2|\ ,\ \pi > {\rm arg} (z_1) > {\rm arg} (z_2) > - \pi\,)
\lb{ggRa}
\ee
where $z_{12} \stackrel{\curvearrowright}{\rightarrow} z_{21} = e^{-i\pi} z_{12}\,$ \cite{FHIOPT}.
It is assumed that ${\hat R}_{12} = P_{12} R_{12}$ (we are using the common tensor product notation)
where $P_{12}\,$ is the permutation matrix, $P^{\a\b}_{~\rho\s} = \d^\a_\s \d^\b_\rho\,,$
and $R_{12}\,$ is a solution of the quantum Yang-Baxter~ equation
\ba
&&R_{12} R_{13} R_{23} = R_{23} R_{13} R_{12}\quad\Leftrightarrow\quad
{\hat R}_{1} {\hat R}_{2} {\hat R}_{1} = {\hat R}_{2} {\hat R}_{1} {\hat R}_{2}\ ,\quad{\hat R}_i := {\hat R}_{i i+1}\nn\\
&&{\rm and,\ trivially,}\quad
{\hat R}_i \, {\hat R}_j = {\hat R}_j \, {\hat R}_i\quad{\rm for}\quad |i-j| > 1\ .
\lb{QYBE}
\ea

The virtue of the exchange relations (\ref{ggRa}) is that they reveal, along with the
left $G$-symmetry (acting on the capital latin indices of $g^A_{~\a}(z)$), also right
{\em quantum group} \cite{D} invariance with respect to transformations satisfying the {\em $RTT$ relations}
\be
R_{12} T_1 T_2 = T_2 T_1 R_{12}\quad\Leftrightarrow\quad {\hat R}_{12} T_1 T_2 = T_1 T_2 {\hat R}_{12}
\lb{RTT}
\ee
which is the quantum counterpart of the Lie-Poisson symmetry of the corresponding classical Poisson brackets.
The relations (\ref{QYBE}) identify ${\hat R}_i$ as generators of the (non-abelian) braid group statistics of the model.

The first sign that the WZNW model is somehow related to quantum groups appeared in \cite{TK}.
Although it became soon clear that the quantum group symmetry does not hold in the unitary version of the model
(in particular, the quantum group representation ring does not close on the "physical" representations),
it seems to be the appropriate internal ("gauge") symmetry for a logarithmic extension of it
(see e.g. \cite{HP, FGST1, FHT7}).

The monodromy matrix $M$ obeys the {\em reflection equation}
\be
M_1\, R_{12}\, M_2\, R_{21} = R_{12}\, M_2\, R_{21}\, M_1 \quad\Leftrightarrow\quad
{\hat R}_{12}\, M_2\, {\hat R}_{12}\, M_2 = M_2\, {\hat R}_{12}\, M_2\, {\hat R}_{12}\ ,
\lb{exM}
\ee
while its exchange relations with $g(z)\,$ read
\ba
&&g_1 (z)\, R_{12}^- M_2 = M_2\, g_1(z)\, R_{12}^+ \qquad (\, R^-_{12} := R_{12}\,,\ R^+_{12} := R_{21}^{-1}\, )\quad\Leftrightarrow\nn\\
&&M_1 \, g_2(z) = g_2(z)\,{\hat R}_{12}\, M_2\, {\hat R}_{12}\ .
\lb{Mgq}
\ea

The quantum group properties of the chiral field $g(z)\,$ become transparent by taking as $R_{12}$
the $U_q ({\cal G}_{\subbbc})$ Drinfeld-Jimbo quantum $R$-matrix (where ${\cal G}_{\subbbc}\,$ is
the complexification of the Lie algebra ${\cal G}\,$ of $G\,$) and performing the {\em factorization} of $M\,$
into a product $M_+ M_-^{-1}$ of two upper, resp. lower~ triangular~ matrices such that
\be
{\rm diag}\, M_+ = {\rm diag}\, M_-^{-1}\ ,\ \
R_{12} M_{\pm 2} M_{\pm 1} = M_{\pm 1} M_{\pm 2} R_{12}\ ,\ \
R_{12} M_{+2} M_{-1} = M_{-1} M_{+2} R_{12}\ .
\lb{exMpm}
\ee
According to a deep result of Faddeev, Reshetikhin and Takhtajan \cite{FRT}, a quotient of the Hopf algebra generated by
the entries of $M_\pm$ and endowed with a coalgebra structure in which the coproduct, counit and antipode are defined as
\be
\Delta ((M_\pm)^\a_{~\b}) = ( M_\pm )^\a_{~\s}\otimes (M_\pm)^\s_{~\b}\ ,\quad
\varepsilon ((M_\pm)^\a_{~\b}) = \d^\a_\b\ ,\quad S ((M_\pm)^\a_{~\b}) =  (M_\pm^{-1})^\a_{~\b}\ ,
\lb{Hopf-FRT}
\ee
respectively, is equivalent to a certain cover~ $U_q$ of $U_q ({\cal G}_{\subbbc})\,.$ The exchange relation
\be
M_{\pm 2}\, g_1(z)\, M_{\pm 2}^{-1}\ \ (\,= M_{\pm 2}\, g_1(z)\, S(M_{\pm})_2 = Ad_{M_{\pm 2}} g_1(z) \, )\ = g_1(z)\, R_{12}^\mp
\lb{gMpm}
\ee
(leading to (\ref{Mgq})) implies that each row of $g(z) = (g^A_{~\a}(z))\,$ is a $U_q$ {\em vector operator}.
The factorization of $M$ actually involves a "quantum prefactor" \cite{FHIOPT};
in particular,~ for $G=SU(n)$ when the deformation parameter~ is~ $q=e^{-i\frac{\pi}{h}}\,,\ h=k+n\,,$
\be
M = q^{\frac{1}{n} - n}\, M_+ M_-^{-1}\qquad\  (\,{\cal G}_{\subbbc} = s\ell(n)\,)\ .
\lb{factorM}
\ee

The quantum $SU(n)$ WZNW monodromy matrix $M$ and its components $M_\pm\,,$ as matrices with {\em non-commutative}
entries, are the main objects of interest for us in this paper.
In Section 2 we remind the FRT construction and provide some important technical details of it.
Section 3 is devoted to the diagonalization of $M\,.$
In the last Section 4 we introduce the quantum determinant $\det_q (M)$ \cite{FH2}
and discuss some of its properties.
The results are illustrated by explicit formulae for small $n\,.$

\section{$U_q$ in disguise: the FRT construction}
\label{sec:2}

One of the amazing results in \cite{FRT} is that a quotient of the $RTT$ algebra (\ref{RTT}), regarded as a deformation
of the algebra of functions on a matrix Lie group $\, G\,,$ is Hopf dual to a certain cover of the
QUEA $U_q({\cal G})\,.$ The "classical" ($q=1$) counterpart of this fact is the realization, due to L. Schwartz,
of the universal enveloping algebra $U({\cal G})$ as the non-commutative algebra of distributions on $G$ supported
by its unit element, $U({\cal G})\simeq C^{-\infty}_e(G)$ (see Theorem 3.7.1 in \cite{C06}).
The details below concern the case ${\cal G} = s\ell (n)\,.$
As shown in \cite{FRT}, the Hopf algebra (\ref{exMpm}), (\ref{Hopf-FRT}) is dual to $Fun(SL_q(n))\,,$
the ${\rm det}_q(T)= 1\,$ quotient of the {\em RTT}\, algebra (\ref{RTT})
(for an appropriate definition of the quantum determinant) with coalgebra relations written in matrix form as
\be
\Delta (1) = 1\otimes 1\ ,\quad \Delta (T) = T\otimes T\ ,\quad \varepsilon (T) = \id\ ,\quad S(T) = T^{-1}\ .
\lb{coRTT}
\ee
The Chevalley generators of $U_q (s\ell(n))$ obey the commutation relations
\ba
&&K_i K_j = K_j K_i\ ,\quad K_i \, E_j \, K_i^{-1} = q^{c_{ij}} \, E_j \ , \quad
K_i \, F_j \, K_i^{-1} = q^{-c_{ij}} \, F_j \ ,\nn\\
&&[E_i , F_j] = \delta_{ij} \, \frac{K_i - K_i^{-1}}{q-q^{-1}} \ , \qquad i,j = 1,\ldots ,n-1
\lb{CRq}
\ea
and, for $n>2\,,$ also the {\it $q$-Serre relations}
\ba
&&E_i^2 \, E_j + E_j \, E_i^2 = [2]\, E_i \, E_j \, E_i \ , \qquad
F_i^2\, F_j + F_j \, F_i^2 = [2]\, F_i \, F_j\, F_i  \nn\\
&&{\rm for} \quad \vert i-j\vert = 1\ ,\qquad
[E_i , E_j] = 0 = [F_i , F_j] \quad \mbox{for} \quad \vert i-j \vert > 1 \ .\qquad
\lb{Sq}
\ea
Here $(c_{ij})$ is the $s\ell (n)$ Cartan matrix, $c_{ii} = 2 \,,\ c_{i\,i\pm 1} = -1 \,,\ c_{ij} = 0\,$ for
$\vert i-j \vert > 1\,.$ The coalgebra structure is defined on the generators as follows:
\ba
&&\Delta (K_i) = K_i \otimes K_i \ , \ \Delta (E_i) = E_i \otimes K_i + \id \otimes E_i\ ,\
\Delta (F_i) = F_i \otimes \id + K_i^{-1} \otimes F_i \ ,
\lb{coalg}\\
&&\varepsilon (K_i) = 1 \,, \ \varepsilon (E_i) = \varepsilon (F_i) = 0 \,,\quad
S(K_i) = K_i^{-1}\,,\  S(E_i) = -E_i K_i^{-1}\,,\ S(F_i) = -K_i F_i\ .
\nn
\ea
On the other hand, using the explicit form of the Drinfeld-Jimbo $U_q(s\ell(n))\ R$-matrix,
\be
R_{12} = ( R^{\a\b}_{~\rho\s} )\ ,\ \
R^{\a\b}_{~\rho\s} = q^\frac{1}{n}\left( \d^\a_{\rho}\d^\b_{\s} + (q^{-1}- q^{\epsilon_{\a\b}} )
\d^\a_{\s}\d^\b_{\rho} \right)\ ,\ \
\epsilon_{\a\b} =
\left\{
\begin{array}{ll}
\, {~~}1&, \quad \a > \b\\
\, {~~}0&, \quad\a = \b\\
\, - 1&,\quad \a < \b
\end{array}
\right.\, ,
\lb{R}
\ee
Eqs. (\ref{exMpm}) give rise to the following relations for~ the components of $M_\pm$:
\ba
&&[ (M_\pm)^\a_{~\rho} , (M_\pm)^\b_{~\s} ] =
(q^{\epsilon_{\s\rho}} - q^{\epsilon_{\a\b}} )\, (M_\pm)^\a_{~\s} (M_\pm)^\b_{~\rho}\ ,
\lb{RM}\\
&&[ (M_-)^\a_{~\rho} , (M_+)^\b_{~\s} ] = (q^{-1} - q^{\epsilon_{\a\b}})\, (M_+)^\a_{~\s} (M_-)^\b_{~\rho} -
(q^{-1} - q^{\epsilon_{\s\rho}})\, (M_-)^\a_{~\s} (M_+)^\b_{~\rho}\ .
\nn
\ea
We will denote
\be
{\rm diag}\, M_+  = {\rm diag}\, M_-^{-1}\ =: D = (d_\a\, \d^\a_\b)\ ,\quad \det D := \prod_{\a=1}^n d_\a = 1\ ,
\lb{MpmD1}
\ee
thus introducing a quotient of the algebra (\ref{exMpm}). From (\ref{RM}) we obtain, in particular,
\ba
&&d_\a\, d_\b = d_\b\, d_\a 
\ ,\lb{dMpm}\\
&&d_\a\, (M_+)^\b_{~\a} = q^{-1}\,  (M_+)^\b_{~\a}\, d_\a\ , \quad d_\b (M_+)^\b_{~\a} = q\, (M_+)^\b_{~\a}\, d_\b\ ,
\qquad\ \ \ \,\a > \b\ ,\nn\\
&&d_\a\, (M_-)^\a_{~\b} = q\,  (M_-)^\a_{~\b}\, d_\a\ , \qquad d_\b (M_-)^\a_{~\b} = q^{-1}\, (M_-)^\a_{~\b}\, d_\b\ ,
\qquad \a > \b\ , \nn\\
&&[ (M_-)^\a_{~\b} , (M_+)^\b_{~\a} ] = \l\, ( d_\a^{-1} d_\b  - d_\a d_\b^{-1} )\ ,\qquad \a > \b\qquad (\,\l
= q-q^{-1}\,)\ .
\nn
\ea
As $d_\a$ commute, their order in the product defining $\,\det D$ in (\ref{MpmD1}) is not important.
Using the triangularity of $M_+$ and $M_-$ in deriving (\ref{dMpm}) is crucial.
Moreover, due to it, the coproduct (\ref{Hopf-FRT}) of a matrix element of $M_+$ or $M_-$
belonging to the corresponding "$m$-th diagonal" (for $m=1,\dots ,n$) contains exactly $m$ summands.
Thus, the diagonal elements $d_\a\,,\ \a=1,2,\dots ,n\,$ ($m=1$) are {\em group-like}
($\Delta (d_\a) = d_\a\otimes d_\a\,,\ \varepsilon (d_\a)=1\,,\ S(d_\a) = d_\a^{-1}$), while
\ba
&&\Delta ((M_+)^i_{~i+1}) = d_i\otimes (M_+)^i_{~i+1} + (M_+)^i_{~i+1} \otimes d_{i+1}\ ,\nn\\
&&\Delta ((M_-)^{i+1}_{~i}) = (M_-)^{i+1}_{~i}\otimes d_i^{-1} + d_{i+1}^{-1} \otimes (M_-)^{i+1}_{~i}
\lb{DeltaMpm}
\ea
for $1\le i\le n-1$ (here $m=2$). The comparison with (\ref{coalg}) suggests that
\be
(M_+)^i_{~i+1} = x_i\, F_i\, d_{i+1}\ ,\quad (M_-)^{i+1}_{~i} = y_i\, d_{i+1}^{-1}\, E_i\ , \quad d_i^{-1} d_{i+1} = K_i
\lb{MpmFE}
\ee
where $x_i$ and $y_i$ are some yet unknown $q$-dependent coefficients. For $\a=i+1\,,\ \b = i\,,$
the second and third relation in (\ref{dMpm}) as well as the condition (\ref{MpmD1}) are satisfied if
\be
d_\a = k_{\a-1} k^{-1}_\a\qquad (\,k_0 = k_n = 1\,)\ ,
\lb{dkk}
\ee
the new set of independent Cartan generators $k_1, \dots , k_{n-1}$ obeying
\ba
&&k_i = \prod_{\ell =1}^i d_\ell^{-1}\ ,\quad K_i = k_{i-1}^{-1} k_i^2 k_{i+1}^{-1}\ ,\quad i=1,2,\dots,n-1\ ,\nn\\
&&k_i k_j = k_j k_i\ ,\quad k_i\, E_j = q^{\d_{ij}} E_j\, k_i\ ,\quad k_i\, F_j = q^{-\d_{ij}} F_j\, k_i\ ,\nn\\
&&\Delta (k_i) = k_i\otimes k_i\ ,\quad \varepsilon (k_i)=1\ ,\quad S(k_i) = k_i^{-1}\ .\lb{dk}
\ea
Inserting (\ref{MpmFE}) into the last Eq.(\ref{dMpm}) and using the second and third relation (\ref{dMpm})
from which it follows that $[ d_{i+1} , (M_-)^{i+1}_{~i} (M_+)^i_{~i+1} ] = 0$, we obtain
\be
x_i\, y_i = - \l^2\ ,\qquad i=1,\dots , n-1\ .
\lb{xiyi}
\ee
The commutation relation (\ref{RM}) of $(M_+)^i_{~i+2}$ with $d_\a$ (\ref{dkk})
suggests that $(M_+)^i_{~i+2}$ contains the step operators $F_i$ and $F_{i+1}\,$ only. Assuming that it is proportional to
$(F_{i+1} F_i - z\, F_i\, F_{i+1} ) D_{i+2}$ where $D_{i+2}$ is some group-like element
and $z$ is another unknown $q$-dependent coefficient,
taking the corresponding coproduct (\ref{Hopf-FRT}) and using (\ref{MpmFE}), (\ref{coalg}), we obtain
\be
(M_+)^i_{~i+2} = - \frac{x_i x_{i+1}}{\l}\,[F_{i+1},  F_i ]_q\, d_{i+2}\ ,\quad (\, [ A , B ]_q := A B - q B A )\ .
\lb{M+i2}
\ee
A similar calculation shows that
$
(M_-)^{i+2}_{~i} = \frac{y_i y_{i+1}}{\l}\, d_{i+2}^{-1}\,[ E_i , E_{i+1} ]_{q^{-1}}\ .
$
We will fix the coefficients $x_i$ and $y_i$ satisfying (\ref{xiyi}) in a symmetric way:
$
x_i = - \l \,,\ y_i = \l \ .
$
The commutators
\ba
&&[ (M_+)^i_{~i+1} , (M_+)^i_{~i+2} ]_q = 0\ ,\quad\ [(M_+)^i_{~i+2} , (M_+)^{i+1}_{~i+2} ]_q = 0\ ,\nn\\
&&[ (M_-)^{i+1}_{~i} , (M_-)^{i+2}_{~i} ]_q = 0\ , \qquad [ (M_-)^{i+2}_{~i} , (M_-)^{i+2}_{~i+1} ]_q = 0
\lb{MpmMpmq}
\ea
are in fact the non-trivial $q$-Serre relations (\ref{Sq}) written as
\ba
&&[ F_i , [ F_i , F_{i+1} ]_{q^{-1}} ]_q \, = \, 0  = \  [ F_{i+1} , [ F_{i+1} , F_i ]_q \, ]_{q^{-1}} \ ,\nn\\
&&[ E_i , [ E_i , E_{i+1} ]_{q^{-1}} ]_q = \, 0 \, = \, [ E_{i+1} , [ E_{i+1} , E_i ]_q \, ]_{q^{-1}} \ .
\lb{Sq-alt}
\ea
One can obtain in a similar way the higher off-diagonal terms of the matrices $M_\pm\,$
(for example, $(M_+)^1_{~4} = - \l\, [ F_3 , [ F_2 , F_1 ]_q ]_q\, d_4$). The result can be summarized in
\be
M_+ = (\id - \l\, N_+ ) \, D\ ,\qquad M_- = D^{-1} \, (\id + \l\, N_- )
\lb{MpmNpmD}
\ee
where the {\em nilpotent} matrices $N_+$ and $N_-$ are upper and lower triangular, respectively, with matrix
elements given by the corresponding (lowering and raising) {\em Cartan-Weyl} generators,
while the non-trivial entries $\, d_\a\,,\ \a=1,\dots ,n\,$ of the diagonal matrix $D$
are expressed in terms of $k_i$ (\ref{dkk}). Writing $K_i = q^{H_i}\,,\ i=1,\dots , n-1$ and using (\ref{dk})
allows to present $k_i$ as $k_i = q^{h^i}$ where $h^i$ are dual to the fundamental weights,
\be
H_i = \sum_{j=1}^{n-1} c_{ij}\, h^j = 2\, h^i - h^{i-1} - h^{i+1}\ .
\lb{Hh}
\ee
As $\,\det\, c^{(n)} = n\,$ for $\, c^{(n)} := (c_{ij})^{s\ell(n)}\,,$ (\ref{Hh}) infers that
an inverse formula expressing $k_i$ in terms of $K_i$ would involve "$n$-th roots" of the latter\footnote
{The determinant of the $s\ell(n)\,$ Cartan matrix obeys
$$
\det\, c^{(n)} = 2\,\det\, c^{(n-1)} - \det\, c^{(n-2)}\ ,\quad \det\, c^{(2)} = 2\ ,\quad \det\, c^{(3)} = 3
\quad\Rightarrow\quad \det\, c^{(n)} = n\ .
$$}; indeed,
\be
h^i = \sum_{j=1}^{n-1} (c^{-1})^{ij} H_j =
\sum_{j=1}^i j\, ( 1-\frac{i}{n} )\, H_j + \sum_{j=i+1}^{n-1} i\,( 1 - \frac{j}{n})\, H_j\ .
\lb{hH}
\ee
Thus the Hopf algebra $U_q\,$ generated by $E_i , F_i , k_i\,$ is an {\em $n$-fold cover} of $U_q(s\ell (n))\,.$

Note that the $U_q$ invariance of the vacuum vector can be written as
\be
X \vac = \varepsilon (X) \vac\qquad \forall X\in U_q\ ,
\lb{Uqvac}
\ee
where $\varepsilon (X)$ is the counit (see (\ref{Hopf-FRT}) or, equivalently, (\ref{MpmNpmD}), (\ref{coalg}), (\ref{dk})).

\smallskip

We display below the matrices $D$ and $N_\pm$ (\ref{MpmNpmD}) in the cases $n=2$ and $n=3\,.$

\smallskip

{$\bf n=2$:}
\be
D = \left(\matrix{k^{-1}&0\cr 0&k}\right)\quad (\, K = k^2\,) \ \ ,\quad
N_+ = \left(\matrix{0&F\cr0&0}\right)\ ,\quad  N_- = \left(\matrix{0&0\cr E&0}\right)\ ,
\lb{MD2}
\ee

{$\bf n=3$:}
\ba
&&D = \left(\matrix{k_1^{-1}&0&0\cr 0&k_1 k_2^{-1}&0\cr 0&0& k_2}\right)\qquad\quad
(\, K_1 = k_1^2 k_2^{-1}\ ,\ \ K_2 = k_1^{-1} k_2^2 \, ) \ ,\nn\\
&&N_+ = \left(\matrix{0&F_1&[F_2 , F_1 ]_q\cr0&0&F_2\cr 0&0&0}\right)\ ,\qquad
N_- = \left(\matrix{0&0&0\cr E_1&0&0\cr [E_1 , E_2 ]_{q^{-1}}& E_2&0}\right)\ ,\lb{MD3}\\
&&(\id + \l N_- )^{-1} = \id - \l \left(\matrix{0&0&0\cr E_1&0&0\cr [E_1 , E_2 ]_q& E_2&0}\right)\ .
\lb{MD3inv}
\ea

\section{The diagonal monodromy matrix $M_p$}
\label{sec:3}

The natural solution of the diagonalization problem for the chiral $SU(n)$ WZNW monodromy matrix $M$
appears to be the diagonal matrix $M_p\,$ defined as
\be
M_p \,\, a\, =\, a\, M\ ,\quad M_p = q^{1-\frac{1}{n}}\,{\rm diag}\, (q^{-2 p_1}\,,\dots , q^{-2 p_n})
\lb{Mpa=aM}
\ee
(see e.g. \cite{FHIOPT}). Here $q^{p_i}$ form a commutative set of operators ($q^{p_i} q^{pj} = q^{p_j} q^{p_i}$) satisfying
$\prod_{i=1}^n q^{p_i} = 1\,,$ the {\em zero modes'} matrix (with non-commutative entries) $\, a$ obeys the relations
\be
q^{p_j}\, a_{\alpha}^i = a^i_\a\,\, q^{p_j + \d^i_j - \frac{1}{n}} \ ,\qquad
{\hat R}_{12} (p)\, a_1\, a_2\, =\, a_1\, a_2\, {\hat R}_{12}
\lb{QMA}
\ee
as well as an appropriate ($n$-linear) determinant condition, and ${\hat R}_{12} (p)$ in (\ref{QMA})
is a solution of the quantum {\em dynamical} Yang-Baxter~ equation \cite{HIOPT}.

The $q^{1-\frac{1}{n}}$ prefactor of $M_p$~ (\ref{Mpa=aM}) has a quantum origin \cite{FHIOPT, FH2}.
Applying both sides of the first relation (\ref{Mpa=aM}) to the vacuum and using
(\ref{factorM}), (\ref{Uqvac}) and the first equation (\ref{QMA}), we deduce that the equality
\be
a^i_\a\,\, q^{-2p_i}\, {\mid 0 \rangle} = q^{1-n}\, a^i_\a\, {\mid 0 \rangle}
\lb{aqp-vac}
\ee
should hold for any $i$ (and $\a$). The natural way to satisfy (\ref{aqp-vac}) is to set
\be
q^{p_i}\,{\mid 0 \rangle} = q^{\frac{n+1}{2} - i}\,{\mid 0 \rangle} \ ,\quad i=1,\dots, n\ , \qquad
a^i_\a \,{\mid 0 \rangle} = 0\quad{\rm for}\quad i\ge 2\ .
\lb{qpan}
\ee
Here $p^{(0)}_i = \frac{n+1}{2} - i$ are the "barycentric coordinates" ($\sum_{i=1}^n p^{(0)}_i = 0$) of the Weyl vector
$\rho$ in the orthogonal basis of the $s\ell(n)$ weights.

\smallskip

These two relations give rise to a Fock representation of the zero modes' matrix algebra
generated by polynomials  ${\cal P}(a)\,$ applied to the vacuum vector. For homogeneous polynomials,
the action of $a^i_\a$ on the vector ${\cal P}(a)\,{\mid 0 \rangle}$ can be depicted as adding a box to the $i$-th row of a Young-type diagram.
In the case of admissible $~s\ell (n)$ diagrams (associated to irreducible representations (IR) with highest weight $\L$)
the eigenvalues of $q^{p_i}$ on ${\cal P}^\L (a)\,{\mid 0 \rangle}$ are expressed in terms of the
barycentric coordinates of the {\em shifted} weight $\L + \rho\,.$ For $q$ generic,
the Fock space is in fact a {\em model space} (a direct sum of all IR with multiplicity one) of $U_q\,$ \cite{FHIOPT}.
In the case at hand $q$ is an (even) root of unity, and a more complicated structure including indecomposable
$U_q$ representations occurs (see \cite{FHT7} where the simplest, $n=2$ case has been studied).

\smallskip

The first equation (\ref{QMA}) implies the following exchange relation of $M_p$ and $a$:
\be
M_{p1} a_2 = q^{- 2 \s_{12}} a_2 M_{p1}\ \Leftrightarrow\
a_1 M_{p2}\, a_1^{-1} = q^{2 \s_{12}} M_{p2}\ ,\quad
(q^{2 \s_{12}})^{ij}_{\ell m} = q^{2(\d_{ij}-\frac{1}{n})}\, \d^i_\ell\, \d^j_m\ .
\lb{aMp}
\ee
On the other hand, the exchange relation between $M$ and $a$ is similar to (\ref{Mgq}):
\be
a_1\,R_{12}^-\, M_2\,=\, M_2\, a_1\,R_{12}^+\qquad\Leftrightarrow\qquad
M_1 \, a_2 = a_2\, {\hat R}_{12} \, M_2\, {\hat R}_{12}\ .
\lb{aM}
\ee
The compatibility of Eqs. (\ref{aMp}) and (\ref{aM}) requires the relation
\be
{\hat R}^{-1}_{12}(p)\,=\, q^{2 \s_{12}} \,M_{p2}\, {\hat R}_{12}(p)\, M_{p 1}^{-1}
\lb{RpHIOPT}
\ee
to hold (it takes place indeed, being equivalent to Eq.(6.17) of \cite{HIOPT}
with ${\hat R}_{12}(p) \leftrightarrow {\hat R}^{-1}_{12}(p)$).
To prove this, we start with (\ref{aM}) and then use $M = a^{-1} M_p\, a$ (\ref{Mpa=aM}), the second equation
(\ref{QMA}) rewritten as $a_2\, {\hat R}_{12}\, a_2^{-1} = a_1^{-1} \, {\hat R}_{12}(p) \, a_1\,,$ and (\ref{aMp}):
\ba
&&M_1 \, a_2 = a_2\, {\hat R}_{12} \, M_2\, {\hat R}_{12}\quad\Rightarrow\quad
(a_1^{-1} \, M_{p1}\, a_1)\, a_2 = a_2\, {\hat R}_{12}\, (a_2^{-1}\, M_{p2}\, a_2)\, {\hat R}_{12}\quad\Rightarrow\nn\\
&&a_1^{-1} \, M_{p1}\, a_1\, = (a_2\, {\hat R}_{12}\, a_2^{-1})\, M_{p2}\, (a_2\, {\hat R}_{12}\, a_2^{-1} )
\quad\Rightarrow\lb{long-calc}\\
&&a_1^{-1} \, M_{p1}\, a_1\, = (a_1^{-1} \, ({\hat R}_{12}(p) \, a_1 ) \, M_{p2}\, (a_1^{-1} \, {\hat R}_{12}(p)) \, a_1)
\quad\Rightarrow\nn\\
&&M_{p1} = {\hat R}_{12}(p) \, (a_1  \, M_{p2}\, a_1^{-1}) \, {\hat R}_{12}(p)
\quad\Rightarrow\quad{\hat R}_{12}^{-1}(p)\, = \,q^{2 \s_{12}} M_{p2}\, {\hat R}_{12}(p)\, M^{-1}_{p1} \ .
\nn
\ea
It is easy to verify Eq.(\ref{RpHIOPT})~ for~ $n=2$ when
\ba
&&\hat R^{\pm 1}_{12} (p) = q^{\pm\frac{1}{2}}\,\left(\matrix{q^{\mp 1}&0&0&0\cr 0&\frac{q^{\mp p}}{[p]}&q^{-\a}\frac{[p-1]}{[p]}&0\cr
0&q^{\a}\frac{[p+1]}{[p]}&-\frac{q^{\pm p}}{[p]}&0\cr 0&0&0&q^{\mp 1}}\right)\ ,\quad
M_p = q^{\frac{1}{2}}\,\left(\matrix{q^{-p}&0\cr 0&q^p}\right)\qquad\quad
\lb{RpMpn2}
\ea
(here $p := p_{12}$ and $\a = \a(p)$),~ so that
\ba
&&q^{-\frac{1}{2}} M_{p2} = {\rm diag}\, (q^{-p}\,,\, q^p\,,\, q^{-p}\,,\, q^p)\ ,\quad
q^{\frac{1}{2}} M^{-1}_{p1} = {\rm diag}\, (q^{p}\,,\, q^p\,,\, q^{-p}\,,\, q^{-p})\ ,\nn\\
&&q^{2\s_{12}} = {\rm diag}\, (q\,,\, q^{-1}\,,\, q^{-1}\,,\, q )\ .
\lb{diagM-q2s}
\ea

\section{The quantum determinant $\det_q (M)$}
\label{sec:4}

As shown in \cite{FH2}, the appropriate definition of the quantum determinant of $M$ is
\be
{\det}_q (M) := \frac{1}{[n]!}\,\e_{\a_1 \dots \a_n}\,
\left[ ({\hat R}_{12} {\hat R}_{23} \dots {\hat R}_{n-1\, n} M_n )^n\right]^{\a_1 \dots \a_n}_{~\b_1 \dots \b_n}\,\e^{\b_1\dots \b_n}\ .
\lb{detM}
\ee
Here $[n]! = [n] [n-1] \dots [1]$ and the {\em quantum} antisymmetric tensors
vanish whenever some of their indices coincide, while their non-zero components are given by
\be
\label{q-eps}
\varepsilon^{\alpha_1 \ldots \alpha_n} = \varepsilon_{\alpha_1 \ldots \alpha_n} = q^{- \frac{n(n-1)}{4}} \,
(-q)^{\ell (\a )}\quad \Rightarrow\quad\varepsilon_{\alpha_1 \ldots \alpha_n}
\varepsilon^{\alpha_1 \ldots \alpha_n} = [n]!
\ee
for $(\a_1 , \dots , \a_n )$ a permutation of $( n , \dots , 1)\,$ of length $\ell(\a)\,.$

The corresponding independent definition of ${\det}_q (M_\pm )$ does not involve the $R$-matrix and is thus simpler;
due to the triangularity of the matrices, only the $n!$ products of (commuting) diagonal entries survive in the sum
so that, by (\ref{q-eps}), the end result complies with (\ref{MpmD1}):
\be
{\det}_q (M_\pm) :=
\frac{1}{[n]!}\,\e_{\a_1 \dots \a_n}\,(M_\pm )^{\a_n}_{~\b_n}\dots (M_\pm )^{\a_1}_{~\b_1} \,\e^{\b_1\dots \b_n}
= \prod_{\a =1}^n (M_\pm )^{\a}_{~\a} = \prod_{\a =1}^n d_\a^{\pm 1} = 1\ .
\lb{detMpmvar1}
\ee

One can prove that the formula (\ref{detM}) possesses the following factorization property.
Substituting $M$ by (\ref{factorM}) (including the prefactor!),
one obtains just the product of the quantum determinants of $M_+$ and $M_-^{-1}$ (both equal to $1$), and hence
\be
{\det}_q (M) = {\det}_q (M_+) \,.\, {\det}_q (M_-^{-1})\, = 1\ .
\lb{MMMpm}
\ee
Of course, this is a highly desirable result, as it appears as a quantum counterpart of the similar classical property.

\smallskip

We will end up by calculating $\det_q (M)$ for $n=2$ directly from (\ref{detM}).
In this case $\e_{12} = \e^{12} = - q^{\frac{1}{2}}\,,\ \e_{21} = \e^{21} = q^{-\frac{1}{2}}\,,$ and with
\be
{\hat R}_{12} = q^{\frac{1}{2}}\,\left(\matrix{ q^{-1}&0&0&0\cr 0&-\l&1&0\cr 0&1&0&0\cr 0&0&0&q^{-1}}\right)\ ,\quad
M := \left(\matrix{m^1_{~1}&m^1_{~2}\cr m^2_{~1}&m^2_{~2}} \right)
\lb{RMn2}
\ee
we obtain the expression
\ba
&&{\det}_q (M) = \frac{1}{[2]}\,\e_{\a\b} \left( {\hat R}_{12} M_2 {\hat R}_{12} M_2 \right)^{\a\b}_{~\rho\s} \e^{\rho\s} =\nn\\
&&= \frac{q^2}{[2]}\, (m^1_{~1} m^2_{~2} +\, m^2_{~2} m^1_{~1} +
q\,\l\,m^2_{~2} - q^{-2} m^1_{~2}m^2_{~1} - m^2_{~1}m^1_{~2} )\qquad
\lb{DqMn2}
\ea
which reproduces the classical one,~
$\,m^1_{~1} m^2_{~2} - m^1_{~2} m^2_{~1}\,,~$ for $q=1$ and commuting $m^\a_{~\b}$.
Through (\ref{MpmNpmD}) and (\ref{MD2}), the entries of $M= q^{-\frac{3}{2}}\,M_+\,M_-^{-1}$ are expressed in terms of
the $U_q$ generators:
\be
m^1_{~1} = q^{-\frac{1}{2}} (\l^2 FE + q^{-1}K^{-1})\,,\ \
m^1_{~2} = -q^{-\frac{1}{2}} \l FK \,,\ \ m^2_{~1} = -q^{-\frac{1}{2}} \l E\,,\ \
m^2_{~2} = q^{-\frac{1}{2}} K\ .
\lb{Mab2}
\ee
(Note that only $k^2 = K \in U_q(s\ell(2))$ appears in (\ref{Mab2}) and not $k\in U_q$ alone \cite{FGST1, FHT7}.)
Now using $K E = q^2 E K\,,\ [E,F]=\frac{K-K^{-1}}{\l}\,,\ [2]=q+q^{-1}$ we obtain
\be
{\det}_q (M) = \frac{1}{[2]}\,(2\, q^{-1}  - \l^2\, [E,F] K + \l\, K^2 ) = 1\ ,
\lb{detqMn=2}
\ee
as prescribed by (\ref{MMMpm}).

\smallskip

\begin{acknowledgement}
L.H. thanks the organizers of the 9th International Workshop
"Lie Theory and Its Applications in Physics" (LT-9), 20-26 June 2011, held in Varna, Bulgaria.

The work of L.H. has been supported in part by the Bulgarian National Science Fund (grant DO 02-257) and
INFN, Sezione di Trieste.
P.F. acknowledges the support of the Italian Ministry of University and Research (MIUR).

\end{acknowledgement}

\biblstarthook{}

\end{document}